\documentclass[]{rsos}

\usepackage{bm}
\usepackage{graphicx}
\graphicspath{{./}{figs/}}

\begin{document}

\title{The counterbend dynamics of cross-linked filament bundles and flagella}

\author{
Rachel Coy$^{1}$, Hermes Gad\^elha$^{2}$}

\address{$^{1}$CoMPLEX, University College London, London WC1E 6BT, UK.\\
$^{2}$Department of Mathematics, University of York, York YO10 SDD, UK.\\}

\subject{xxxxx, xxxxx, xxxx}


\corres{Insert corresponding author name\\
\email{hermes.gadelha@york.ac.uk}}

\begin{abstract}
Cross-linked filament bundles, such as in cilia and flagella, are ubiquitous in biology. They are considered in textbooks as simple filaments with larger stiffness. Recent observations of flagellar counterbend, however, show that induction of curvature in one section of a passive flagellum instigates a compensatory counter-curvature elsewhere, exposing the intricate role of the diminutive cross-linking proteins at large-scales. We show that this effect, a material property of the cross-linking mechanics, modifies the bundle dynamics non-trivially, and induces a bimodal $L^2-L^3$ length-dependent material response that departs from the Euler-Bernoulli theory. Hence, the use of simpler theories to analyse experiments can result in paradoxical interpretations. Remarkably, the counterbend dynamics instigates counter-waves in opposition to driven oscillations in distant parts of the bundle, with potential impact on the regulation of flagellar bending waves. These results have a range of physical and biological applications, including the empirical disentanglement of material quantities via counterbend dynamics.
\end{abstract}


\begin{fmtext}
\section{Introduction}
The spontaneous generation of harmonic bending waves along a sperm flagellum has been a source of fascination since it was reported on in the late 17th century \cite{howards_antoine_1997}. It was not until 1968, however, that the fundamental mechanism behind the flagellar wave propagation was unveiled \cite{satir_studies_1968}. ATP-induced inter-microtubule tangential motion is converted into transversal forces that are capable of bending the flagellar assembly altogether, laying the empirical basis for the sliding filament theory for eukaryotic flagellum. 

\end{fmtext}

\maketitle
Notably, almost one decade before the discovery of the interfilament sliding \cite{satir_studies_1968}, the existence of such active elements along the sperm flagellum was theorized via a simple fluid-structure interaction model \cite{machin_wave_1958}. Machin demonstrated that the combined action of viscous and elastic dissipation experienced by a slender filament rapidly damps any driven oscillation along its length, thus requiring the action of contractile elements in order to sustain large waving amplitude \cite{brokaw_flagellar_1972}. Later, Goldstein and co-workers \cite{Wiggins2} elegantly demonstrated that the elastohydrodynamics of any simple Euler-Bernoulli filament moving in a viscous fluid leads to a hyperdiffusive dissipation of bending, characterized by a bending penetration length $\ell_b$, which can be further exploited to extract material parameters from biological filaments in a wide range of length-scales \cite{Wiggins1998a}. Hitherto the elastohydrodynamics of active and passive filaments have generated a vast literature of analytical, computational and empirical studies across disciplines \cite{Gadelha2010,Camalet,Bourdieu95,Goldstein1995,Fu2008,Yu,olson_modeling_2013,Tornberg2003,kantsler_fluctuations_2012,sanchez_spontaneous_2012}.

Despite the inherent complexity of filament bundles \cite{Heussinger2010,Claessens2008,claessens2006actin,fawcett_textbook_1994,Afzelius,tolomeo1997mechanics,minoura1999direct,book:Alberts,howard2001,Lindemann1973,Okuno1980,Okuno1979}, as exemplified by the axonemal flagellum \cite{Afzelius,fawcett_textbook_1994,Afzelius}, with its $9+2$ cross-linked  microtubule doublets arranged in a cylindrical fashion \cite{fawcett_textbook_1994},  the textbook elastic bending stiffness has been estimated using a simplistic linear relation between bending moment and curvature \cite{book:Alberts,howard2001,Lindemann1973,Okuno1980,Okuno1979}, as derived from Euler-Bernoulli rod theory \cite{book:antman}. Incidentally, the inadequacy of classical rod theories, from Euler-Bernoulli to Timoshenko and Cosserat \cite{book:antman},  emerged  via paradoxical counterbend empirical responses, Fig. \ref{fig_schematic}(a), first observed by Lindemann and co-workers \cite{Lindemann2005,Pelle2009}, and later  captured via a geometrically exact mechanical model by Gad\^elha et al \cite{gadelha_counterbend_2013} (Fig. \ref{fig_schematic}(a)). These studies revealed how the induction of curvature in one section of a passive sperm flagellum instigates non-trivial compensatory counter-curvature elsewhere, namely the counterbend phenomenon \cite{Lindemann2005,Pelle2009}. They established the critical role of the diminutive elastic linking-proteins while instigating large-amplitude deformations, inherently coupling  distant parts along the bundle assembly, despite their small slenderness ratio \cite{gadelha_counterbend_2013,Lindemann2005,Pelle2009}. More recently, the counterbend phenomenon was also exploited in order to extract material quantities from \textit{Chlamydomonas} flagella \cite{bayly_counterbend}, despite the relatively short length flagella. The dynamical response of the counterbend phenomenon in passive cross-linked bundles still remains unexplored in the literature.

The discovery of counterbend phenomenon highlighted the current need to reassess both the established material measurements \cite{book:Alberts,howard2001,Lindemann1973,Okuno1980,Okuno1979,tolomeo1997mechanics,minoura1999direct,claessens2006actin,Heussinger2010,Claessens2008,everaers1995}, and the resulting mechanical response, from statics to dynamics, of cross-linked filament bundles \cite{book:Alberts,howard2001,Lindemann1973,Okuno1980,Okuno1979}. The former is crucial in a broad range of biological structures, from the cytoskeleton of eukaryotic cells to cellular division, cross-bridge muscle contraction, and locomotion, via structures like the axoneme. A fundamental challenge, both experimentally and theoretically, is therefore to understand how this complex structure yields bulk material properties and overall cellular  mechanical responses and, ultimately, function. In active bundles, the consequences of using inadequate material parameters, which have been  used for the past 30 years to investigate flagellar waves \cite{brokaw_flagellar_1972,Hines78,Brokaw75,Brokaw85,Brokaw75_2,Machin63,lindemann1994geometric,hines1979bend,rikmenspoel1971contractile,Camalet,Riedel2007,bayly_equations_2014,bayly_analysis_2015,sartori_dynamic_2016}, are still unknown. This is further confronted with an increasing number of, repeatedly contradicting, active control models for the flagellar wave coordination \cite{bayly_equations_2014,bayly_analysis_2015,brokaw_flagellar_1972,Brokaw85,Camalet,hines1979bend,lindemann1994geometric,Riedel2007,sartori_dynamic_2016,Brokaw75,brokaw2014computer,lindemann2010flagellar,brokaw2005computer}. Paradoxically, in order to induce bending waves, flagellar control models rely on the implementation of filament bundle deformations, in distinct material directions,  that are yet to be scrutinized in isolation, from curvature \cite{Brokaw85,sartori_dynamic_2016,Hines78,rikmenspoel1971contractile} to interfilament sliding \cite{brokaw2005computer,brokaw_flagellar_1972,Camalet,hines1979bend,Riedel2007,Brokaw1975}, and axial distortions \cite{bayly_equations_2014,lindemann1994geometric,brokaw2014computer}. This is aggravated by the strong coupling between the unknown activity, and the passive and dissipative components, leading to the non-identifiability of parameters when contrasted against experiments \cite{oriola2017,plouraboue2016identification}. Without the disentanglement between the passive and active elements, and without the rationalization of the resultant mechanical response of cross-linked filament bundles,  it is unclear, for example, which competing flagellar control hypothesis \cite{bayly_equations_2014,bayly_analysis_2015,brokaw_flagellar_1972,Brokaw85,Camalet,hines1979bend,lindemann1994geometric,Riedel2007,sartori_dynamic_2016,Brokaw75,brokaw2014computer,lindemann2010flagellar}, if any, is able to provide a quantitative understanding of the flagellar regulation and, crucially, function of the internal mechanics and structure. Indeed, any comprehensive model of flagellar bending self-organization depends on reliable measurements of mechanical and material properties of the system in absence of activity \cite{gadelha_counterbend_2013,bayly_counterbend}.

Here, we complement the seminal work by Machin \cite{machin_wave_1958} and Goldstein\cite{Wiggins2} on the dynamics of passive filaments, and demonstrate how the nanometric cross-linking proteins that are present in passive cross-linked filament bundles instigate novel dynamical counterbend phenomena. This is in contrast with previous models on flagellar wave coordination \cite{bayly_equations_2014,bayly_analysis_2015,brokaw_flagellar_1972,Brokaw85,Camalet,hines1979bend,lindemann1994geometric,Riedel2007,sartori_dynamic_2016,Brokaw75,brokaw2014computer,lindemann2010flagellar,brokaw2005computer}, which incorporate the cross-linking interaction in conjunction with molecular motor dynamics. We consider the dynamical situation in which only the structural passive elements are present. For axonemal filament-bundles, this corresponds to the empirical situation in which molecular motors are rendered passive \cite{Lindemann2005,Pelle2009,bayly_counterbend}, Fig. \ref{fig_schematic}(a). The filament-bundle elastohydrodynamical model unveils the occurrence of counter-travelling waves in distant parts of bundle, reducing the propulsive potential of driven oscillations, and even reversing the propulsive direction, from pushing to pulling hydrodynamics. We show that the interplay between the interfilament sliding at the base, and cross-linking dissipation elsewhere, give rise to a bimodal $L^2-L^3$ length-dependent material response that departs from canonical Euler-Bernoulli theory. Hence the use of simpler rod theories to analyse experiments can result in paradoxical interpretations. Furthermore, the counterbend dynamics offers a robust way to measure material quantities empirically, bypassing  cumbersome force-displacement experiments at the microscale \cite{Okuno1979,Okuno1980,Lindemann2005,Pelle2009,bayly_counterbend}, Fig. \ref{fig_schematic}(a). These results further suggest that the dynamical counter-wave phenomena is likely to play a critical role on the waveform  organization, and the subsequent wave direction, of long flagella \cite{sartori_dynamic_2016,Hilfinger2009,bayly_analysis_2015}.

\section{Cross-linked filament bundle elastohydrodynamics}

\begin{figure*}[t!]
\centering\includegraphics[width=0.95\textwidth]{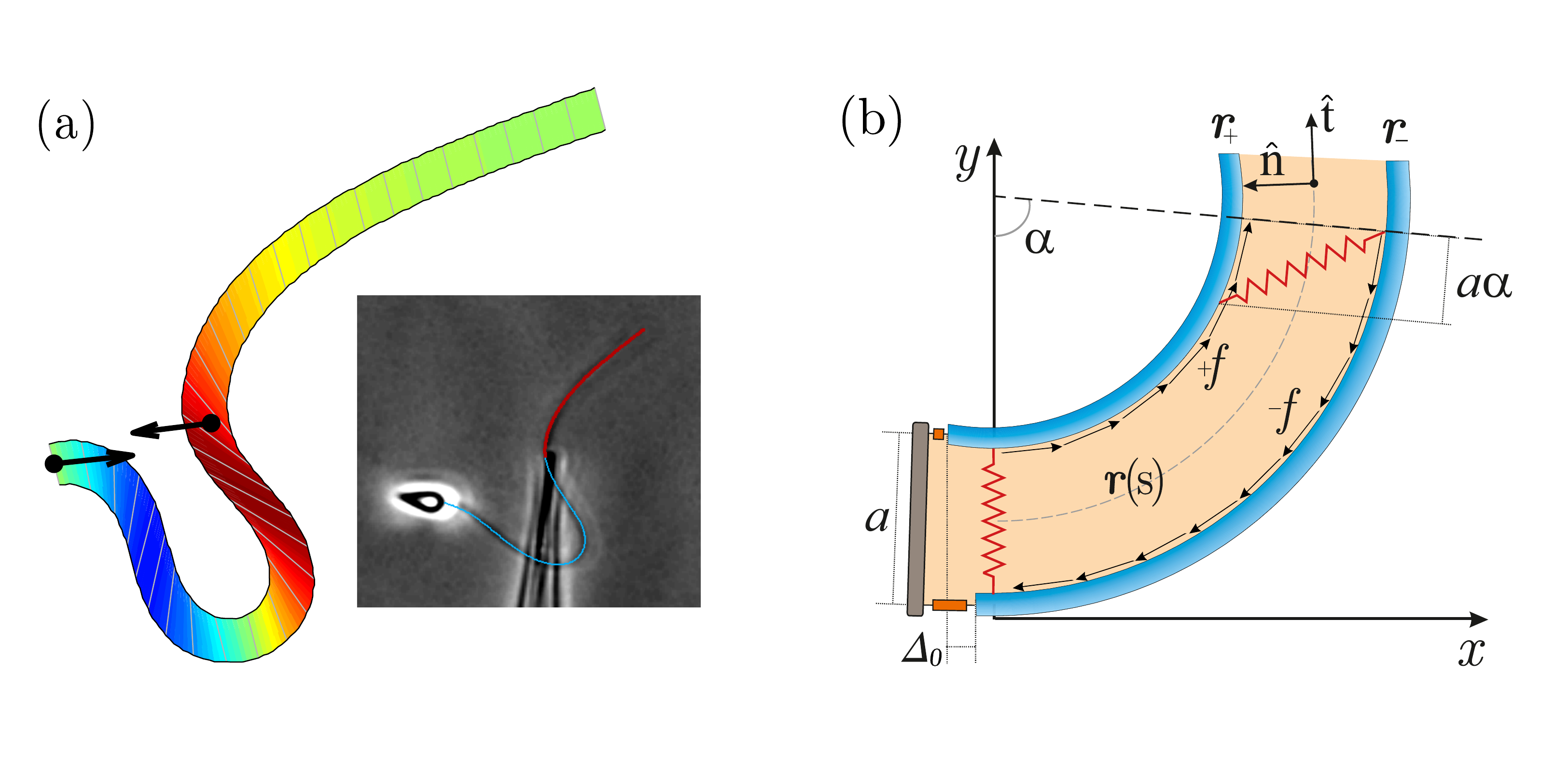}
\caption{\label{fig_schematic} The counterbend phenomenon and geometry of deformation: (a) Micrograph showing  the static configuration of a see urchin sperm rendered passive with its head attached to the coverslip while forced externally by a micro-probe \cite{Pelle2009}, together with the geometrically exact filament-bundle model prediction. Red curve show the model curve fitting result from \cite{gadelha_counterbend_2013}. (b) 2D representation of the axoneme and the sliding filament mechanism with basal compliance \cite{brokaw_flagellar_1972,everaers1995,Riedel2007,gadelha_counterbend_2013}.  Micrograph adapted from Gad\^elha et al. \cite{gadelha_counterbend_2013}.  }
\end{figure*}

We consider a planar representation of cross-linked filament bundles and flagellar axonemes, as depicted in Fig. \ref{fig_schematic}(b), used interchangeably hereafter, composed by two elastic, inextensible filaments that resists deformation with an elastic bending modulus $E$ \cite{brokaw_flagellar_1972,everaers1995, Riedel2007,gadelha_counterbend_2013}.  Each constituent filament   ${\bf r}_\pm(s,t) = {\bf r}(s,t) \pm a/2~\hat{\bf n}(s,t)$ is separated by a distance $a$, much smaller than the filament length  $a \ll L$, normal to the to the centreline ${\bf{r}}(s,t)$ at every point in arclength $s$ and time $t$. Geometry constrains the normal vector $\hat{\bf n}(s,t) = -\sin{\alpha}~{\bf e}_x  + \cos{\alpha}~{\bf e}_y$ to the plane, where $\alpha$ is the angle between the fixed frame $x$-axis and the tangent to the centreline $\hat{\bf{t}} = {\bf r}_s$. Like a rail-track \cite{everaers1995}, the constituent filaments travel distinct contour lengths forcing a geometrical arclength mismatch $\Delta(s,t)=\Delta_0(t) + a(\alpha(s,t) - \alpha_0(t))$, where $\Delta_0$ and $\alpha_0$ are the length mismatch and tangent angle at $s = 0$ (Fig. \ref{fig_schematic})(b). Points of equal contour length along the filament-bundle are connected by elastic links which generate a shearing force,  and thus an internal moment,  proportional to the sliding displacement $f(s,t) = k\Delta(s,t)$ with an elastic sliding resistance $k$. At the basal end, the additional connecting compliance across the filaments, commonly found in spermatozoa and inhomogeneous bundles, is Hookean $\kappa_e\Delta_0(t) = -\int^L_0f(s' ,t)\mathrm{d}s'$ with a spring constant $\kappa_e$ \cite{Riedel2007,gadelha_counterbend_2013} (Fig. \ref{fig_schematic}(b)).

For asymptotically slender filament-bundles, the hydrodynamic forces experienced by an infinitesimal element is anisotropic and linearly related to the local velocity $f_{\text{vis}} = -\zeta_\perp(\hat{\bf n}.{\bf r}_t)\hat{\bf n}-\zeta_\parallel(\hat{\bf t}.{\bf r}_t)\hat{\bf t}$,  where $\zeta_\perp, \zeta_\parallel$ are the lowest order resistive coefficients derived from inertialess hydrodynamics \cite{book:Alberts,howard2001}. Contact forces are not defined constitutively due to the inextensibility constraint \cite{book:antman}. The filament-bundle elastohydrodynamics is governed by the  balance of contact forces and contact moments 
\begin{equation}
\label{eqn:gov}
 -E\alpha_{ssss}+a^2k\alpha_{ss}=\zeta_\perp\alpha_t,
\end{equation}
simplified here for small curvatures~\cite{Wiggins2,Wiggins1998a,Camalet,Riedel2007}. The filament-bundle shape is given by the initial value problem  ${\bm{r}}(s,t) = \bm{r}(0,t) + \int^s_0(\cos{\alpha(s',t)},\sin{\alpha(s',t)})\; \mathrm{d}s'$ for an arclength $s$ and time $t$. Boundary conditions ensure the total balance of forces, $F(s) = -E\alpha_{ss} +af(s,t)$, and torques, $M(s) = -E\alpha_{s} + a\int^L_sf(s',t)\mathrm{d}s$, acting on the bundle \cite{book:antman,gadelha_counterbend_2013}, as detailed in the Supporting Information (SI). The resulting cross-linking mechanics couples distant parts along the bundle  via the total momentum balance,  now modified non-locally by  $f(s,t) =-\gamma ka\int^L_0(\alpha(s')-\alpha_0)\;\mathrm{d}s'+ ka(\alpha(s)-\alpha_0)$ (SI). Dissipation from different material directions are mediated by the hydrodynamic drag. The filament-bundle dynamics is dictated by the interplay between the elastohydrodynamic hyperdiffusion \cite{Wiggins2,Wiggins1998a} and the cross-linking diffusion \cite{brokaw_flagellar_1972,everaers1995,Riedel2007}.  These boundary moments alter the hyperdiffusion balance in Eq. \ref{eqn:gov} non-trivially, and instigates novel long-range phenomena as we explore below.

\begin{figure*}[t!]
\centering\includegraphics[trim=42 0 0 0,clip,width=\textwidth]{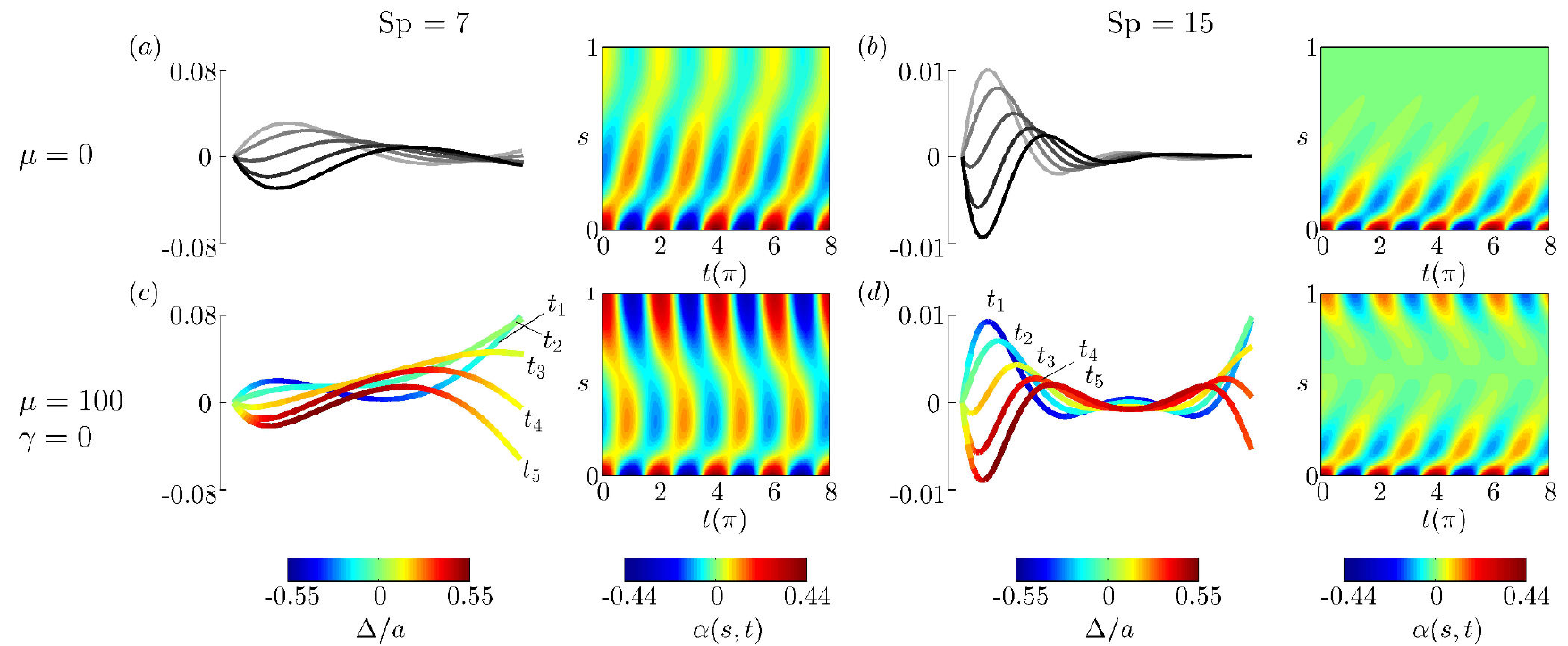}
\caption{\label{fig1}Counter travelling wave formation in filament bundles. (a,b) Euler-Bernoulli hyperdiffusive waveforms and (c,d) filament bundle waveforms for $\gamma=0, \mu = 100$. The rescaled sliding displacement $\Delta/a$ is overlaid in the waveform in (c,d). The time progression of the waveforms runs from $t_1$ to $t_5$, covering one half of the periodic solution. For the $\alpha$ plots, the time is labelled in multiples of $\pi$.}
\end{figure*}

\subsection{The counterbend dynamics: angular actuation}
The post-transient behaviour of the shape dynamics is captured by single frequency solutions of the form $\alpha(s,t) = \text{Re}\{\tilde{\alpha}(s)e^{-i\omega t}\}$. After convenient rescaling, the eigenvalue problem reduces to $r^4-\mu r^2-i\text{Sp}^4=0$, with eigenfunctions $\alpha(s,t) = \text{Re}\{\sum^4_{j=1}C_je^{r_js-i t}\}$ which  coefficients are determined by non-local boundary moments (SI). After convenient rescaling, the eigenvalue problem reduces to $r^4-\mu r^2-i\text{Sp}^4=0$, with eigenfunctions $\alpha(s,t) = \text{Re}\{\sum^4_{j=1}C_je^{r_js-i t}\}$ which  coefficients are determined by non-local boundary moments (SI). The dimensionless filament-bundle compliance parameter, also referred as sperm number, $\mathrm{Sp} = L(\zeta_\perp \omega/E)^{1/4}$, captures the battle between elastic and viscous forces \cite{Wiggins2}, whilst the sliding resistance parameter, $\mu = a^2L^2k/E$, compares bending stiffness with the cross-linking resistance \cite{Riedel2007,gadelha_counterbend_2013}. The basal compliance is given by $\gamma = kL/(kL+\kappa_e)$, and varies from $\gamma = 0$, corresponding to no interfilament sliding at the base, to $\gamma = 1$, for a free basal sliding \cite{Riedel2007,gadelha_counterbend_2013}.

The emergence of the non-local, counterbend dynamics is depicted in Fig.~\ref{fig1} for a sinusoidal angular actuation of the proximal end (Supplementary Movie 1) with ${\bf r}_t(0,t)=0$, and zero force and  torque condition at the distal end (SI). An angular amplitude of $0.4362$ rad is used to limit the maximum radius of curvature to $0.1$. Fig.~\ref{fig1} contrasts Machin's original solutions, Fig.~\ref{fig1}(a,b), with the filament-bundle post-transient dynamics,  Fig.~\ref{fig1}(c,d). Travelling waves originating from the distal end, indicative of the non-local counterbend effect \cite{gadelha_counterbend_2013}, can be clearly seen in Fig.~\ref{fig1}(c,d).  A relatively small sliding displacement $\Delta$ (overlaid colour in Fig.~\ref{fig1}(c,d)), equivalent to only half of the bundle diameter, is capable of deforming the bundle non-locally with the same magnitude of the imposed actuation.

\begin{figure}[t]
\centering\includegraphics[width=0.8\textwidth]{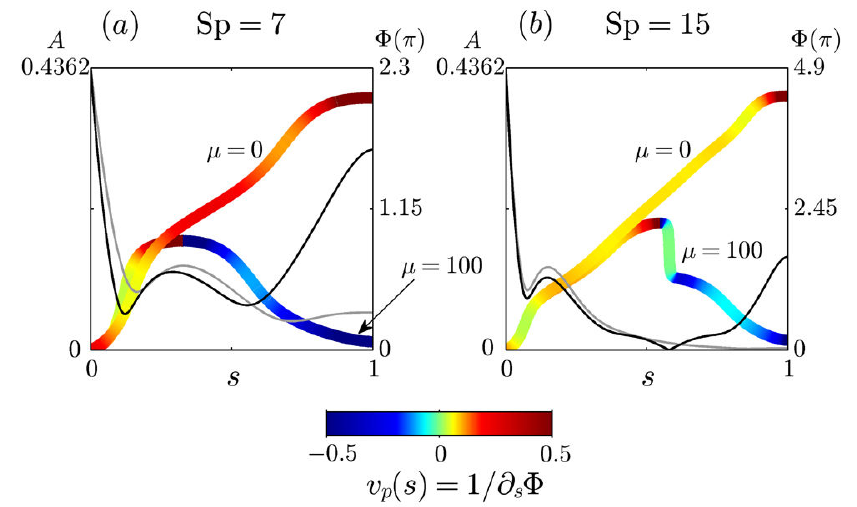}
\caption{\label{velocity plot} The amplitude $A$ of the solutions in the $\mu = 0$ and $\mu = 100$ cases are represented by the grey and black lines respectively; the amplitude of the coloured lines indicates the values of $\Phi$ at each point along the flagella as a multiple of $\pi$, with the colour denoting the velocity. Note that all velocities below $-0.5$ and above $0.5$ are coloured the same as each of these values respectively. For all of the cases featured in the plot, $\gamma = 0$.}
\end{figure}

The amplitude modulation $A(s)$, phase $\Phi(s)$ and velocity of propagation $v_p(s) = 1/\partial_s \Phi$ are depicted in Fig. \ref{velocity plot}, following suitable transformation to solutions of the form $\alpha(s,t) = A(s)\cos{(t-\Phi(s))}$. The abrupt change in wave direction is triggered by the loss in monotonicity of the phase. Non-local counter-waves propagate in opposite direction from the distal end with a non-uniform decaying magnitude along the arclength due to the high-order dissipation. This is in contrast to the one-directional wave propagation of Euler-Bernoulli filaments, $\mu=0$ in Fig \ref{velocity plot}.  The sharp change in wave direction  coincides with reduced wave amplitudes at the point in arclength Fig. \ref{velocity plot}). Interestingly, the non-local actuation of cross-linking moments at distal parts of the bundle is delayed by the overdamped dynamics, as indicated by the proximal-distal phase difference. Higher $\mathrm{Sp}$ causes larger proximal-distal phase mismatch, faster decay of the counterbend wave speed and amplitude, indicative of a destructive interference between proximal and distal waves in Figs.~\ref{fig1}(d) and \ref{velocity plot} (b), demonstrated by the sharp  jump in phase in Fig. \ref{velocity plot}.

\begin{figure}[t]
\centering\includegraphics[width=0.65\textwidth]{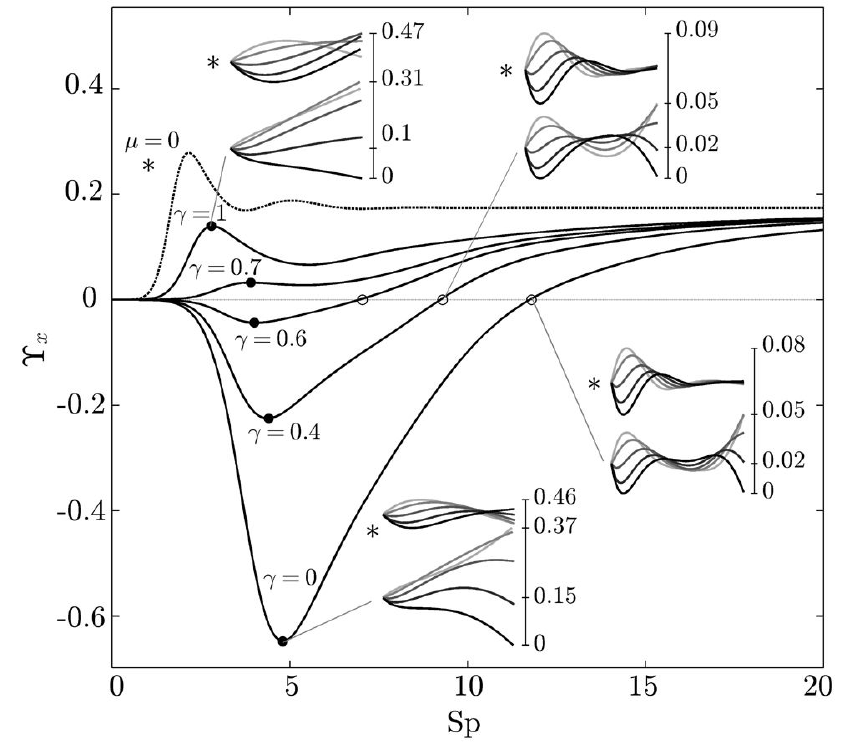}
\caption{\label{force plot}Scaling function $\Upsilon_x$, representative of the propulsive force. All cases plotted use the sliding resistance parameter value $\mu = 100$, except for the dashed line which plots the case $\mu = 0$. Waveforms corresponding to this $\mu = 0$ case are indicated by $*$. The local maxima/minima $\Upsilon_x(\mathrm{Sp}_m)$ in the interval $\mathrm{Sp}_m \in [0,5]$ are labelled by black dots. Positions of the values where $\Upsilon_x = 0$ are labelled by circles. }
\end{figure}

\begin{figure}[h]
\centering\includegraphics[width=0.5\textwidth]{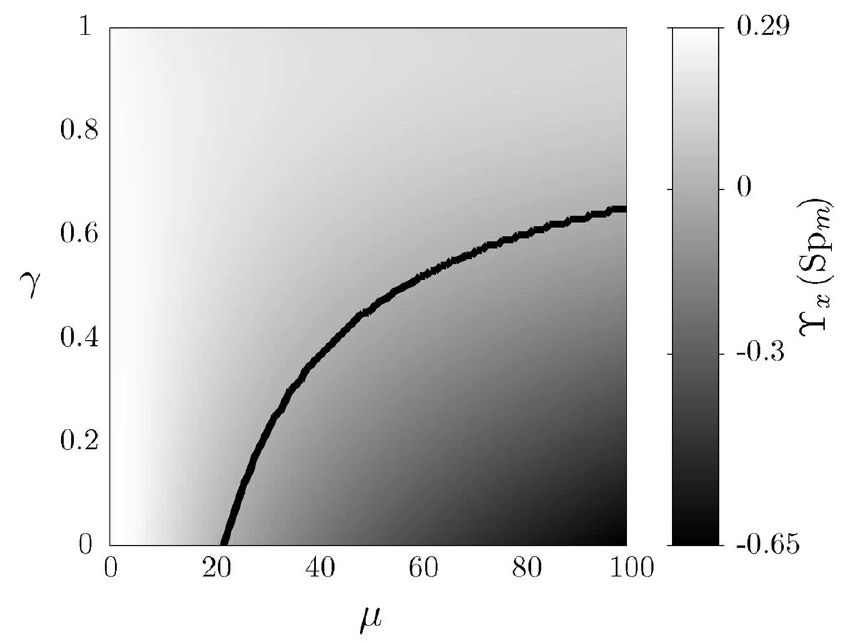}
\caption{\label{maximum force}Local maxima/minima $\Upsilon_x(\mathrm{Sp}_m)$, with $\mathrm{Sp}_m \in [0,5]$, across the parameter space $(\mu, \gamma)$. The black line separates the values for which $\Upsilon_x(\mathrm{Sp}_m)$ is negative from those for which it is positive.}
\end{figure}

The counterbend dynamics impacts significantly the resulting hydrodynamic propulsion. The time-averaged propulsive force $\bar{F}_x$ generated over a period may be written as $$ \bar{F}_x = (\zeta_\perp - \zeta_\parallel)\omega/2\pi\sqrt{E/\omega \zeta_\perp}\Upsilon_x(\mathrm{Sp},\mu,\gamma),$$ where $\Upsilon_x(\mathrm{Sp},\mu,\gamma)$ is the force scaling function, now modified by the cross-linking dynamics, as depicted in Fig. \ref{force plot}. For an effectively stiff Euler-Bernoulli filament ($\mu=0$ and low $\mathrm{Sp}$), no propulsive force can be generated \cite{Wiggins2,Purcell}. This is in accordance with Purcell's scallop theorem where no net propulsion can be achieved via a time reversible motion in inertialess fluids \cite{Wiggins2,Purcell}. As the basal compliance becomes stiffer, by reducing $\gamma$, the cross-linking mechanics switch from a mostly local contribution with small counterbend deformations ($\gamma=1$ in Fig. \ref{force plot}), to a non-local counterbend effect with increasingly large amplitudes at the distal end ($\gamma=0$ in Fig. \ref{force plot}). Ultimately, this causes  the propulsive force to vanish (circles in Fig. \ref{force plot}), and even switch the propulsive direction, thus equivalent to a backward net motion. This is despite the imposed waving direction, which  is counteracted by waves travelling in opposition at distal parts (Fig. \ref{force plot}). The separatrix in Fig.~\ref{maximum force} captures the region in parameter space where the local extrema of $\Upsilon_x(\mathrm{Sp}_m)$ changes sign. Thus this indicates the region where a significant influence of non-local counterbend effect is predicted. This illustrates how the triad  $(\mathrm{Sp},\mu,\gamma)$ may be conveniently tuned to achieve zero, forward or backward propulsion (Fig.~\ref{maximum force}). Reversal in swimming direction may be achieved by simply increasing the frequency of oscillation for instance. Moreover, the cross-linking dissipation does not affect the bundle penetration length $\ell_b=L/\mathrm{Sp}$ \cite{Wiggins2}. Indeed, as $\mathrm{Sp}$ increases, only tangential forces contribute to propulsion, thus the positive asymptote for both Euler-Bernoulli filaments and bundles in Fig. \ref{force plot}.

\subsection*{The bimodal length-dependent relaxation dynamics}
The cross-linking mechanics introduces a diffusion-like time-scale, $L^2\zeta_\perp/a^2k$, with a somewhat weaker $L^2$ geometrical dependence, in addition to the high-order, hyperdiffusive scaling $L^4\zeta_\perp/E$. Indeed, the cross-linking resistance $\mu = a^2L^2k/E$ contrasts the elastohydrodynamic and cross-linking time-scales. The cross-linking resistance depends on geometrical aspects of the bundle, as $\mu$ measures  the ratio between the natural cross-linking elastic length $\ell = \sqrt{E/a^2k}$ relative to total axial length via $\mu=(L/\ell)^2$. The cross-linking elastic length $\ell$ has an important biophysical interpretation: it is the dimensional length by which cross-linking effects become prevalent. The cross-linking mechanics become increasingly important when $L>\ell$, while the opposite is found for $L \sim \mathcal{O}(\ell)$ or smaller. The latter entails the possibility of studying relaxation countebend phenomena by only varying the length of filament bundle, and motivates rescaling $\mu$ relative to the reference length $L_0$, so that $\mu(L) = L^2\mu_0$, $\mu_0 = a^2\text{L}_0^2k/E$ and $L=\text{L}/\text{L}_0$, with smallest dimensionless length $L=1$. 

\begin{figure}[h]
\centering\includegraphics[width=0.5\textwidth]{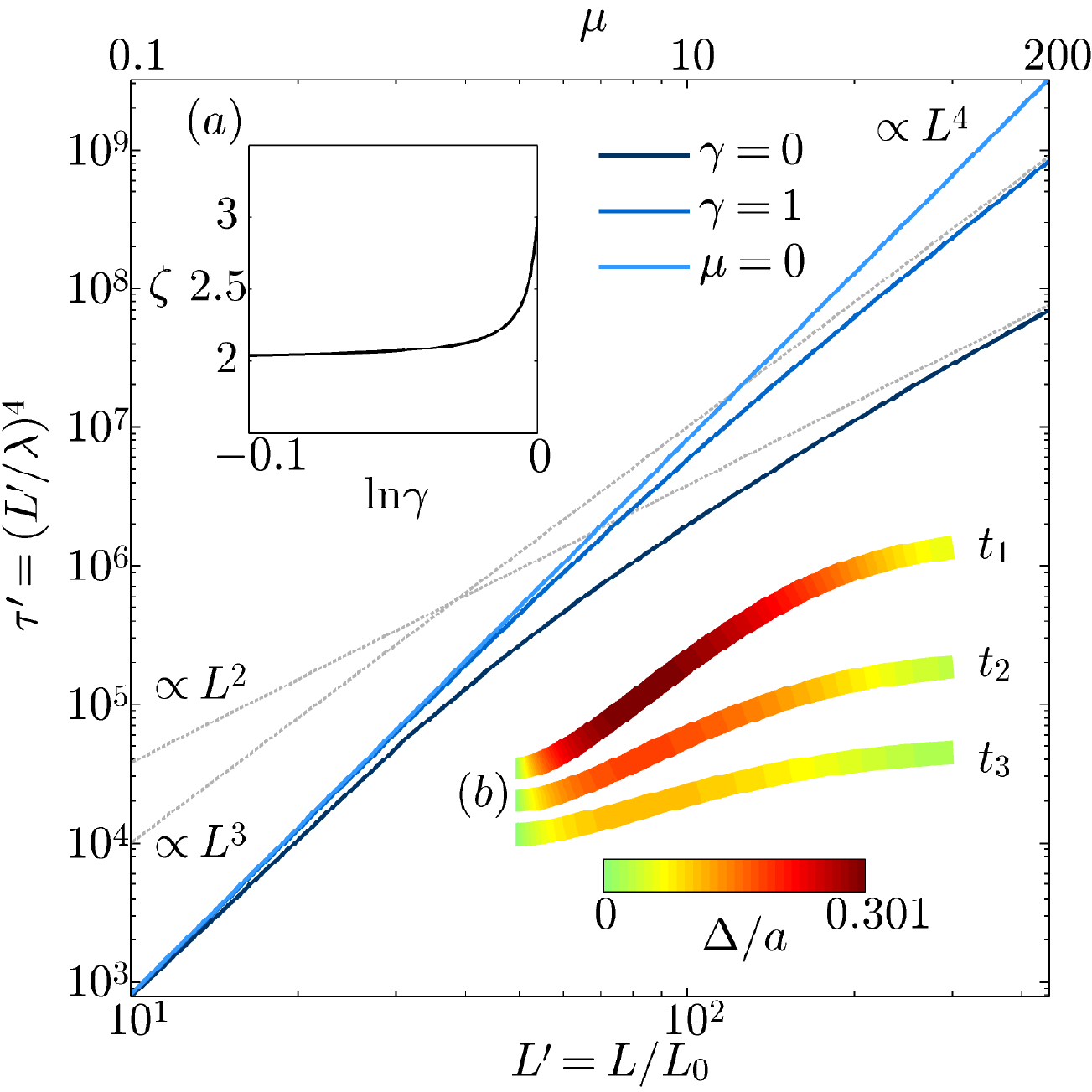}
\caption{\label{relaxationtime}Rescaled relaxation time $\tau'$ as a function of $L'$. The grey lines plotting $L^2$ and $L^3$ are included for comparison. The inset Fig. 5(a) plots the exponent $\zeta$ of the relationship $\tau' \propto L^\zeta$, as calculated for long flagella, against $\ln{\gamma}$; Fig. 5(b) shows the progression of the relaxation of the first mode, with the colour indicating the rescaled sliding displacement $\Delta/a$, for the case $\mu = 100$ and $\gamma = 0$.}
\end{figure}

The non-local, counterbend dynamics  decaying from initial data, with an amplitude $a_n$, is depicted in Fig.~\ref{relaxationtime} for filament-bundles that are clamped at the proximal end (SI). This is characterized by an effective relaxation constant $ \lambda_n^{-4}$ via $\alpha(s,t) = \sum_n a_nS_{n}(s)\,e^{-\lambda_n^{4}t},$  with $S_{n} = C_1\sin({q_{1_n}s})+C_2\cos({q_{1_n}s})+C_3\sinh({q_{2_n} s})+C_4\cosh({q_{2_n}s})$ for a given participating mode $n$. The triad of dissipative contributions acts as an effective dispersion medium for both bending and cross-linking deformations, dictated by the same dispersion relation. However, the mode shape is captured by the wavenumber-eigenvalue coupling $q_{l_n} = \sqrt{ (\sqrt{\mu^2 +4\lambda_n^4} + (-1)^l\mu)/2}$ for $l=1,2$, which is not only influenced by the effective relaxation constant, reminiscent of pure high-order elastohydrodynamic dissipation, but also by the cross-linking diffusion. Boundary conditions define the transcendental solvability condition for $\lambda_n$, which depend implicitly on both $\mu$ and $\gamma$, with an infinite number of mode solutions for each parameter set. Curve-fitting expressions for $\lambda_1$ obtained from numerical solutions are presented in SI.

The non-local cross-linking diffusion introduces a  bimodal length-dependent material response, as illustrated in Fig.~\ref{relaxationtime} for the relaxation time of the fundamental mode 
$$\tau_1(L) = \left(\frac{L}{\lambda_1(L)}\right)^4.$$
The filament-bundle relaxation time departs  from the characteristic $L^4$ dependence of simpler Euler-Bernoulli filaments. Instead, an $L^2$ asymptote arises for long bundles with $\gamma = 0$, while for $\gamma = 1$ the transition is to an $L^3$ behaviour. The length-dependent transition between $L^2$ and $L^3$ modes is governed by the basal compliance (Fig.~\ref{relaxationtime} inset (a)). For asymptotically long filament bundles, the exponent of $\tau_1 \propto L^\zeta$ is quadratic, and remains nearly quadratic until $\gamma$ approaches $1$. Such bijection of the material response entails that simultaneous measurement of the bundle mechanical properties, in different material direction, can be extracted from simple relaxation experiments. In particular, increased inter-filament sliding, concentrated towards the clamped end, induces curvature-reversal for long filament bundles (Fig. \ref{relaxationtime} inset (b)), reminiscent of the counterbend phenomenon \cite{gadelha_counterbend_2013}. Boundary conditions require zero contact forces and torques at the free end, while clamped constraint facilitates the accumulation of cross-linking sliding towards the proximal end. Both bending and cross-linking deformations relax towards the reference configuration with the same effective rate $\lambda_1$ (Fig. \ref{relaxationtime} inset (b)).

\section*{Discussion}
We studied the transient and post-transient  dynamics of overdamped filament bundles that are interconnected by linking elastic proteins. Deformations in distinct material directions, arising from the cross-linking interfilament sliding and pure bending deformation, are coupled with local slender-body hydrodynamics. This leads to an effective dispersion mechanism governed by the superposition of short and long-range dissipation mechanisms. Cross-linking stresses are transmitted to distant parts of the bundle via boundary  balance of moments. The cumulative moments are able to surpass the high-order elastohydrodynamic dissipation, and  shape the bundle structure  non-locally, with increased  influence for long filament-bundles, or equivalently, large $\mu$. 

The delicate interplay between the interfilament sliding at the base and the rest of the bundle results in a bimodal dynamic response, which departs  from the classical Euler-Bernoulli theory \cite{machin_wave_1958,Wiggins2,book:antman}. When the basal sliding is permitted, cross-linking diffusion is mostly local, and acts to effectively reinforce the bundle structure.  Long-range curvature-reversal events, however, are magnified when the basal sliding  is constrained \cite{gadelha_counterbend_2013}. The counterbend dynamics generate spontaneous travelling waves in opposition to driven oscillations, which are capable of suppressing the propulsive potential, and even reverting the direction of propulsion (Fig.~\ref{force plot}). Curvature perturbations diffuse more rapidly, a hundred times faster than Euler-Bernoulli hyperdiffusion with an equivalently higher bending rigidity (Fig. \ref{relaxationtime}). Relatively small cross-linking deformations, up to only $30\%$ of the bundle diameter, are capable of exciting large counterbend modes (Fig. \ref{relaxationtime} inset (b)), and induce a bimodal $L^2-L^3$ length-dependent deviation from the $L^4$-dependence of canonical filaments. Paradoxical measurements may arise if standard  Euler-Bernoulli theory is used to interpret experiments \cite{Lindemann1973,Okuno1979,Lindemann2005,Pelle2009}, as exemplified by the paradoxical length-dependent bending stiffness in microtubules \cite{pampaloni2006thermal}. Indeed, the length-deviation predicted here may be mistakenly interpreted as an effective length-dependent bending rigidity via $L^4\zeta_\perp/E (L)$ \cite{Wiggins2}, if rather the Euler-Bernoulli theory is used; de facto, the Euler-Bernoulli theory is traditionally used since the first measurements of flagellar bundles \cite{Lindemann1973,Okuno1979,book:Alberts}. 

Static, force-displacement experiments that are often used to probe flagellar material quantities \cite{Lindemann1973,Okuno1979,Pelle2009,bayly_counterbend} are cumbersome, see Fig. \ref{fig_schematic}(a). They require  high-precision force calibrated probes and micro-manipulators, and often rely on the rare attachment of the filament's tip to the cover-slip to micro-probe actuation \cite{Lindemann1973,Okuno1979,Pelle2009}. This proximity of the filament bundle to the cover-slip can interfere the interfilament sliding due to surface adhesion, biasing in this way force and shape measurements \cite{gadelha_counterbend_2013}. The counterbend dynamics provides a simpler and robust empirical route for the disentanglement of material parameters. This includes measurements of the basal interfilament elasticity, despite being  deeply embedded at the connecting piece of the bundle (Fig. \ref{fig_schematic}). As a result, standard  microfluidic designs may be explored to induce shape changes dynamically \cite{kantsler_fluctuations_2012}. Likewise, the dynamical counter-wave phenomenon may also inspire the design of artificial swimmers \cite{gadelha2013optimal} that are able to reverse the swimming direction by simply increasing, for instance, the frequency of oscillation (see Fig. \ref{force plot}).

The counter-wave phenomena becomes increasingly important for bundles longer than $\ell = \sqrt{E/a^2k}$, typically $5 \mu m$ for flagella \cite{satir_studies_1968,brokaw_flagellar_1972,wan2016coordinated,Riedel2007,gaffney,sartori_dynamic_2016}. Interestingly, the majority of eukaryotic flagella exceed $\ell$ by few orders of magnitude, from approximately $30\mu m$ for \textit{Chlamydomonas} and sea urchin sperm to almost $200\mu m$ for quail sperm \cite{gaffney,wan2016coordinated}. Cross-linking effects may also become increasingly important during flagellar growth, and  influencing in this way the wave coordination during flagellar reconstitution in \textit{Chlamydomonas}\cite{polin}. Molecular motors organization thus may operate differently for $L<\ell$ and $L>\ell$. Indeed, local flagellar control models \cite{sartori_dynamic_2016,brokaw_flagellar_1972} recently gained empirical support when tested against short flagella experiments \cite{sartori_dynamic_2016}, a regime where counterbend phenomenon may be negligible ($L<\ell$). This is despite of the well-known negative support of curvature control models \cite{Brokaw85,Riedel2007}, tested instead against long flagella ($L>\ell$). The recurrent contradictions between sliding and curvature control models \cite{brokaw_flagellar_1972,Brokaw85,brokaw2005computer,Camalet,Riedel2007,sartori_dynamic_2016} may suggest the occurrence of distinct length-dependent flagellar regimes.

Linear models coupling the molecular motor reaction kinetics with interfilament sliding spontaneously propagate waves along the axoneme via a Hopf bifurcation  \cite{bayly_analysis_2015,Brokaw75_2,Camalet,hines1979bend,Riedel2007}. The resulting wave train is observed to move from tip to base, i.e. in the direction opposite to what one would expect from a local dissipation theory (when the interfilament sliding is prevented at the base) \cite{Camalet,Hilfinger2009,sartori_dynamic_2016}. Incidentally, the basal compliance was observed to influence the direction of wave propagation in flagella self-organizations models \cite{Hilfinger2009,Riedel2007}, demonstrating the sensitivity of the direction of the travelling wave to details of the connecting piece (basal part) and boundary conditions. This is in agreement with the bimodal response predicted here (see Fig. \ref{fig1}), in which counterbend is maximized when $\gamma=0$ \cite{gadelha_counterbend_2013}. The wave direction is influenced non-locally by cross-linking effects whose magnitude is regulated by the basal interfilament sliding (see Figs. \ref{fig1} and \ref{relaxationtime}). This might explain the surprising significance of the basal compliance during the flagellar wave coordination observed recently in empirical studies \cite{wan2016coordinated}, and even flagellar synchronization that may arise without recurring to hydrodynamical coupling \cite{Tam2015}. Nevertheless, the mechanisms by which the flagellar wave direction is selected is poorly understood. Previous studies were reduced to the linear level, and at the nonlinear
 regime, the dynamical instability generates unstable traveling waves that can propagate in both directions, with potencial for multi-frequency modes \cite{oriola2017,oriola2014subharmonic}. The boundary conditions and basal mechanics assist the mode selection nonlinearly, and thus the direction of propagation, emphasizing how the flagellar dynamics is critically dependent on the underlying structural mechanics of the axoneme.  Non-local hydrodynamic interactions \cite{gaffney}, transversal axonemal deformations \cite{bayly_equations_2014,lindemann1994geometric}, and geometrical non-linearities \cite{Gadelha2010,Hines78} are also likely to affect the emergence of self-organization in flagellar systems.

The high-order diffusive interaction, intrinsic to elastohydrodynamic systems in Eq. \ref{eqn:gov}, is observed throughout nature. In non-dilute systems, particles are affected by density variations beyond the nearest neighbours via, for example,  biharmonic interactions  $u_t = D_1\nabla^2u -D_2\nabla^4 u$ \cite{book_murray2001,cohen1981generalized}, thus closely related to Eq.~\ref{eqn:gov}. Despite the relatively short-range influence, such higher-order diffusion instigates non-trivial spatio-temporal dynamics and self-organization in all fields of science \cite{book_peletier2012,book_synchronization}. They drive instabilities and even mediate the coexistence of spatial patterns and temporal chaos, as observed in Kuramoto-Sivashinsky systems \cite{book_synchronization}. Other exemplars of local, higher-order diffusion are found in Ginzburg-Landau superconductors, spatial patterning  in Cahn-Hilliard and biochemical  systems,  plus generalized Fisher-Kolmogorov models, water waves and continuum mechanics systems, among others \cite{book_murray2001,book_peletier2012,book_synchronization}.

In contrast with canonical high-order diffusion systems \cite{book_murray2001,book_peletier2012,book_synchronization}, the flagellar scaffold, or equivalently, any cross-linked filament bundle immersed in a viscous fluid is governed by high-order dispersion medium that is inherently \textit{non-local}. Here, the biharmonic diffusion arises instead via a \textit{local} elastohydrodynamic dissipation \cite{machin_wave_1958,Wiggins2}, while the Fickian-like interaction arises through the \textit{long-range} coupling reflecting the bundle mechanics \cite{brokaw_flagellar_1972,everaers1995,Camalet,gadelha_counterbend_2013},  effectively connecting distant parts of the system via boundary bending moments. This unveils the potential for rich long-range phenomena via reaction-diffusion interactions~\cite{Camalet,Hilfinger2009,oriola2017},  from  non-local  pattern formation to long-range synchronization of auto-oscillators, that are yet to be fully explored in the realm of mathematical biology.

We hope that these results will inspire theoreticians and experimentalists to study the dynamical effects of the conterbend phenomenon in filament-bundle as found throughout nature, including prospects for counterbend reaction-diffusion systems in flagellar dynamics, effectively bridging, non-locally, dynamical systems and PDE's.

\section*{Acknowledgment}
R.C. thanks Cambridge Bridgwater Summer Research Programme. H.G. acknowledges support by the Hooke Fellowship, University of Oxford, and WYNG Fellowship, Trinity Hall, Cambridge. The authors also thank Dr E.A. Gaffney for enlightening discussions. We dedicate this work in memory of Prof. John R. Blake, whose work and devotion will continue to inspire future generations of scientists.



\bibliographystyle{unsrt}
\bibliography{refs}
\end{document}